\documentclass[reprint,twocolumn,aps,prb,superscriptaddress,a4paper,showpacs,showkeys,lengthcheck]{revtex4-1}

\usepackage{graphicx}
\usepackage{type1cm}
\usepackage[normalem]{ulem}

\usepackage{bm,dcolumn,amssymb,amsmath,braket}

\usepackage[dvips]{color}

\newcommand{\new}[1]{ #1 }

\newcommand{\mm}[1]{\mbox{$#1$}}
\newcommand{\dstd}{\mathrm{d}}
\newcommand{\ms}{\mbox{$\mu_{\mathrm{spin}}$}}
\newcommand{\mo}{\mbox{$\mu_{\mathrm{orb}}$}}
\newcommand{\mB}{\mbox{$\mu_{B}$}}

\newcommand{\MM}{\mbox{$\bm{M}$}}
\newcommand{\EF}{\mbox{$E_{F}$}}
\newcommand{\mca}{magnetocrystalline anisotropy}

\newcommand{\sopt}{second order perturbation theory}

\begin{document}

\title{Illustrative view on the magnetocrystalline anisotropy of
  adatoms and monolayers}



\author{O. \surname{\v{S}ipr}} 
\email{sipr@fzu.cz}
\homepage{http://www.fzu.cz/~sipr} \affiliation{Institute of Physics
  ASCR v.~v.~i., Cukrovarnick\'{a}~10, CZ-162~53~Prague, Czech
  Republic }

\author{S. \surname{Mankovsky}} \affiliation{Universit\"{a}t
  M\"{u}nchen, Department Chemie, Butenandtstr.~5-13,
  D-81377~M\"{u}nchen, Germany}

\author{S. \surname{Polesya}} \affiliation{Universit\"{a}t
  M\"{u}nchen, Department Chemie, Butenandtstr.~5-13,
  D-81377~M\"{u}nchen, Germany}

\author{S. \surname{Bornemann}} \affiliation{Universit\"{a}t
  M\"{u}nchen, Department Chemie, Butenandtstr.~5-13,
  D-81377~M\"{u}nchen, Germany}

\author{J. \surname{Min\'{a}r}} \affiliation{Universit\"{a}t
  M\"{u}nchen, Department Chemie, Butenandtstr.~5-13,
  D-81377~M\"{u}nchen, Germany} \affiliation{New Technologies Research
  Centre, University of West Bohemia, Pilsen, Czech Republic}

\author{H. \surname{Ebert}} \affiliation{Universit\"{a}t M\"{u}nchen,
  Department Chemie, Butenandtstr.~5-13, D-81377~M\"{u}nchen, Germany}

\date{\today}

\begin{abstract}
Even though it has been known for decades that the magnetocrystalline
anisotropy is linked to the spin-orbit coupling (SOC), the mechanism
how it arises for specific systems is still subject of debate.  We
focused on finding markers of SOC in the density of states (DOS) and
on employing them for understanding the source of \mca\ for the case
of adatoms and monolayers.  Fully relativistic {\em ab-initio}
KKR-Green function calculations were performed for Fe, Co, and Ni
adatoms and monolayers on Au(111) to investigate changes in the
orbital-resolved DOS due to a rotation of magnetization.  In this way
one can see that a significant contribution to the \mca\ for adatoms
comes from pushing of the SOC-split states above or below the Fermi
level.  As a result of this, the \mca\ energy crucially depends on the
position of the energy bands of the adatom with respect to the Fermi
level of the substrate. This view is supported by model crystal field
Hamiltonian calculations.
\end{abstract}



\maketitle


\section{Introduction}   \label{sec-intro}

Magnetic anisotropy, i.e., the preference of a system for being
magnetized in a certain direction, is one of the key properties
underlying practical use of magnetic materials.  One contribution to
the magnetic anisotropy comes from the classical interaction of magnetic
dipoles.  This mechanism stands behind the so-called shape anisotropy
and can be described using classical physics --- although a
quantum-mechanical description has been developed as
well.\cite{Jan88a,BMB+12} Another contribution, which becomes
important in particular for small systems such as atomic clusters or
nanostructures, comes from the spin-orbit coupling (SOC).  This {\em
  magnetocrystalline anisotropy} can only be described within a
relativistic quantum-mechanical formalism.  We will deal exclusively
with this SOC-induced contribution in this work.

A quantitative measure of the magnetocrystalline anisotropy is the
magnetocrystalline anisotropy energy (MAE), i.e., the difference
between total energies of the system for two orientations of the
magnetization $\bm{M}$.  Evaluating the MAE is often numerically very
difficult because one has to subtract two large numbers to get a small
difference between them.  To get accurate results, one has to tune
several technical parameters such as integration mesh setup in 
$\bm{k}$-space\cite{RKF+01,JSW+03} or the adequacy of the basis set.  For
supported nanostructures, the treatment of the substrate is also very
important.\cite{SBME10,SBE+14} A lot of attention was devoted to these
issues recently.

Nevertheless, there is also another line of research on the \mca, namely,
the effort to understand its mechanism intuitively and, in particular,
to see which electronic structure features participate in the
phenomenon.  One possibility is to use perturbation theory and to
describe spin-orbit interaction approximately within the two-component
formalism by the SOC term $H_{\text{SOC}}=\xi\bm{L}\cdot\bm{S}$, where
$L$ and $S$ are the orbital and spin angular momentum operators and
$\xi$ is the SOC strength.  For systems studied here the lowest-order
non-vanishing contribution to the total energy is the second-order
term,
\begin{equation}
\Delta E^{(2)} \: = \: - 
\sum_{i \in \text{occ} \atop j \in \text{unocc}} 
\frac{ \left| \bra{\psi_{i}} H_{\text{SOC}} \ket{\psi_{j}} \right|^{2} }
     { E_{j} \, - \, E_{i} }
\;\; .
\label{2nd}
\end{equation}
Relying on \sopt\ has led to concepts such as scaling of the MAE with
the square of the SOC strength or the frequently used Bruno and van
der Laan formulae relating the MAE to the anisotropy of the orbital
magnetic moment.\cite{Bru89,vdL98,ABS+06,KMI+14,KS+15} On the other hand, as
the sum in Eq.~(\ref{2nd}) involves a large number of summands which
may be of comparable magnitude, it may be very difficult to identify
just a few terms as the dominant ones and in this way to link MAE of a
particular system to specific features in the electronic structure.
Getting a simple intuitive understanding of the MAE by looking on the
interaction between individual states thus may be very hard to achieve
--- despite the effort and interesting results
obtained.\cite{WWF93a,LMH+97,GC+12} Approaches that focus on integral
quantities such as a corresponding susceptibility (still within 
\sopt) could have a more general use.\cite{KS+15}

However, other mechanisms of generating the \mca, not accounted for by
\sopt, are also possible and were discussed in the past.  In
particular, Eq.~(\ref{2nd}) cannot be used if degenerate levels are
coupled.  For that situation another mechanism contributing to the MAE
was suggested, namely, a SOC-induced splitting of states that would be
degenerate otherwise.\cite{WWF93b,DKS+94,MHB+96,RKF+01,KS+15} If some
of these states are pushed above or below the Fermi level $E_{F}$, a
large change of the total energy occurs.  For layered and bulk systems
this effect may not be dominant because relevant states occupy only a
restricted region in $\bm{k}$-space.\cite{MHB+96,LMH+97,Blu+99}
However, the situation could be different for adatoms and clusters,
where there is no dispersion in $\bm{k}$-space.

The question then remains whether there exist in reality systems where
the origin of \mca\ can be traced to a SOC-induced splitting of
otherwise degenerate states at $E_{F}$ and where this mechanism can be
effectively visualized in terms of integral quantities such as the
density of states (DOS).  Such a mechanism could give rise to a large
MAE. In fact lately there have been intensive efforts to understand
how the MAE could be made as large as
possible.\cite{SCM+07,AKA+14,RBR+14,KW+14,BDS+15} A better intuitive
insight into the \mca\ beyond the perturbation theory might be useful
in this context.  From a more general point of view, it is desirable
to have a framework that would enable to visualize the emergence of
the \mca\ by means of simple concepts.

We decided therefore to perform a detailed {\em ab initio}, i.e.,
material-specific study of \mca\ for Fe, Co, and Ni adatoms and
monolayers on Au(111).  The motivation for this choice is that only
little hybridization between 3$d$ states and Au states is
expected.\cite{BSM+12} For adatoms, the 3$d$ states could thus have an
atomic-like character where the effect of SOC-induced splitting of
states should be larger than for delocalized states.  Comparison
between adatoms and monolayers could further elucidate the role of
different factors.  We employ a fully relativistic framework (solving
the four component Dirac equation) to treat the SOC as accurately as
possible.  The application of Green function formalism allows a proper
treatment of adatoms, avoiding possible artefacts that might arise
from a supercell approach.

Our paper is organized as follows. First we introduce our
computational method and the investigated systems.  Then we present
numerical values of MAE and magnetic moments.  The main emphasis is
put on showing how the SOC affects the DOS resolved into components
according to the magnetic quantum numbers. We demonstrate that the
effect of SOC is much larger if the magnetization is perpendicular to
the plane than if it is in-plane.  This effect is reproduced using a
simple crystal-field Hamiltonian.  Some technical details related to
projecting the DOS onto magnetic quantum number components for a
magnetic system are described in the appendix.


\section{Methods}

\subsection{Computational scheme}   \label{sec-comput}

The electronic structure is calculated within the {\em ab initio} spin
density functional framework, relying on the local spin density
approximation with the Vosko, Wilk and Nusair parametrization for the
exchange and correlation potential.\cite{VWN80} The electronic
structure is described, including all relativistic effects, by the
Dirac equation, which is solved using the spin-polarized relativistic
multiple-scattering or Korringa-Kohn-Rostoker (KKR) Green function
formalism \cite{EKM11} as implemented in the {\sc sprkkr}
code.\cite{sprkkr-code} The potential was treated within the atomic
sphere approximation (ASA).  For the multipole expansion of the Green
function, the angular momentum cutoff \mm{\ell_{\mathrm{max}}}=3 was
used.  The energy integrals were evaluated by contour integration on a
semicircular path within the complex energy plane using a Gaussian
mesh of 32 points.  The integration over the $\mathbf{k}$ points was
done on a regular mesh, using 10000 points in the full surface
Brillouin zone.

This work deals with adatoms and monolayers on a substrate.  The Green
function formalism allows to treat the substrate as truly
semi-infinite: the electronic structure is relaxed within the topmost
seven substrate layers while at the bottom of this relaxation zone the
electronic structure is matched to the bulk via the decimation
technique.  Monolayers are dealt with in the same manner as the clean
substrate, just adding a layer of 3$d$ atoms on top.  The vacuum is
represented by four layers of empty spheres.  Adatoms are treated as
embedded impurities: first one calculates the electronic structure of
a semi-infinite host and then solves the Dyson equation for an
embedded impurity cluster.\cite{MBS+06} The impurity clusters we used
contain 62 sites in total; this includes one 3$d$ adatom, 25~substrate
atoms and the rest are empty spheres.

The MAE is calculated as a difference of total energies for
\mm{\bm{M}\| \bm{\hat{x}}}\ and \mm{\bm{M}\| \bm{\hat{z}}},
\begin{equation}
E_{\text{MCA}} \: = \: E^{(x)} \, - \, E^{(z)}
\;\; .
\end{equation}
Accordingly, a positive MAE means that the easy axis of magnetization
is out-of-plane.

If the Dirac equation is used, the influence of SOC cannot be isolated
in a straightforward way. One can achieve it, nevertheless, using an
approximate two-component scheme \cite{EFVG96} where the SOC-related
term is identified via relying on a set of approximate radial Dirac
equations.  This scheme was used in the past to investigate the
influence of SOC on various properties including the MAE.\cite{SBE+14}
In this work we use this scheme to suppress the SOC when investigating
the DOS in Sec.~\ref{sec-dos}.  If SOC is to be included, the DOS can
be calculated either using the full Dirac equation or using the
approximative scheme;\cite{EFVG96} the corresponding curves in the
graphs agree within the thickness of the line, demonstrating that both
schemes are equivalent as concerns the DOS.  On the other hand, there
are small yet identifiable differences between both schemes concerning
the MAE (about 10~\% in case of adatoms and about 20~\% in case of
monolayers).  The results presented in Sec.~\ref{sec-moms} were
obtained using the fully relativistic scheme.


\subsection{Investigated systems}   \label{sec-systems} 
 
\begin{figure}
\includegraphics[viewport=0 0 540 520,height=30mm]{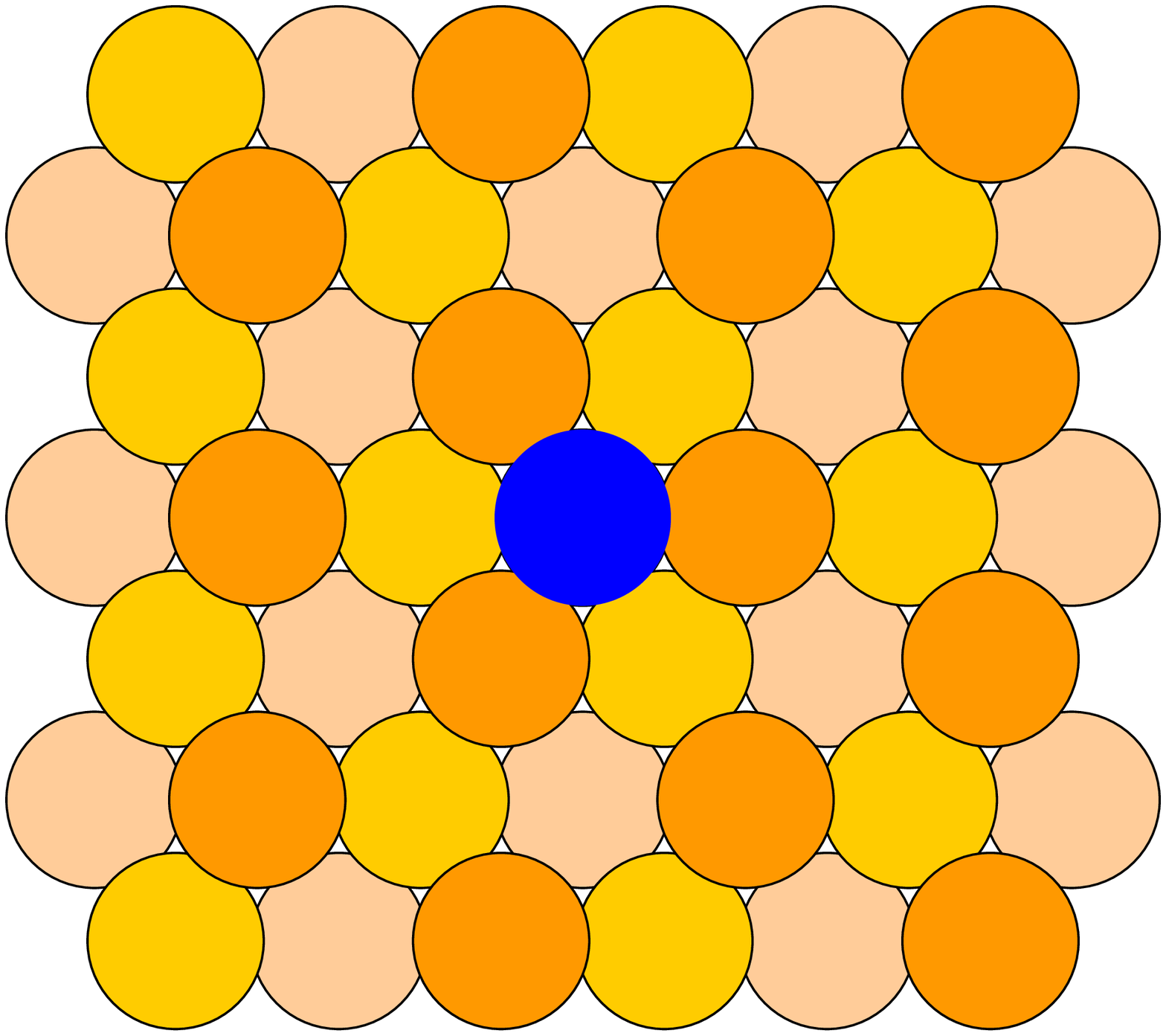}%
\includegraphics[viewport=-160 0 540 520,height=30mm]{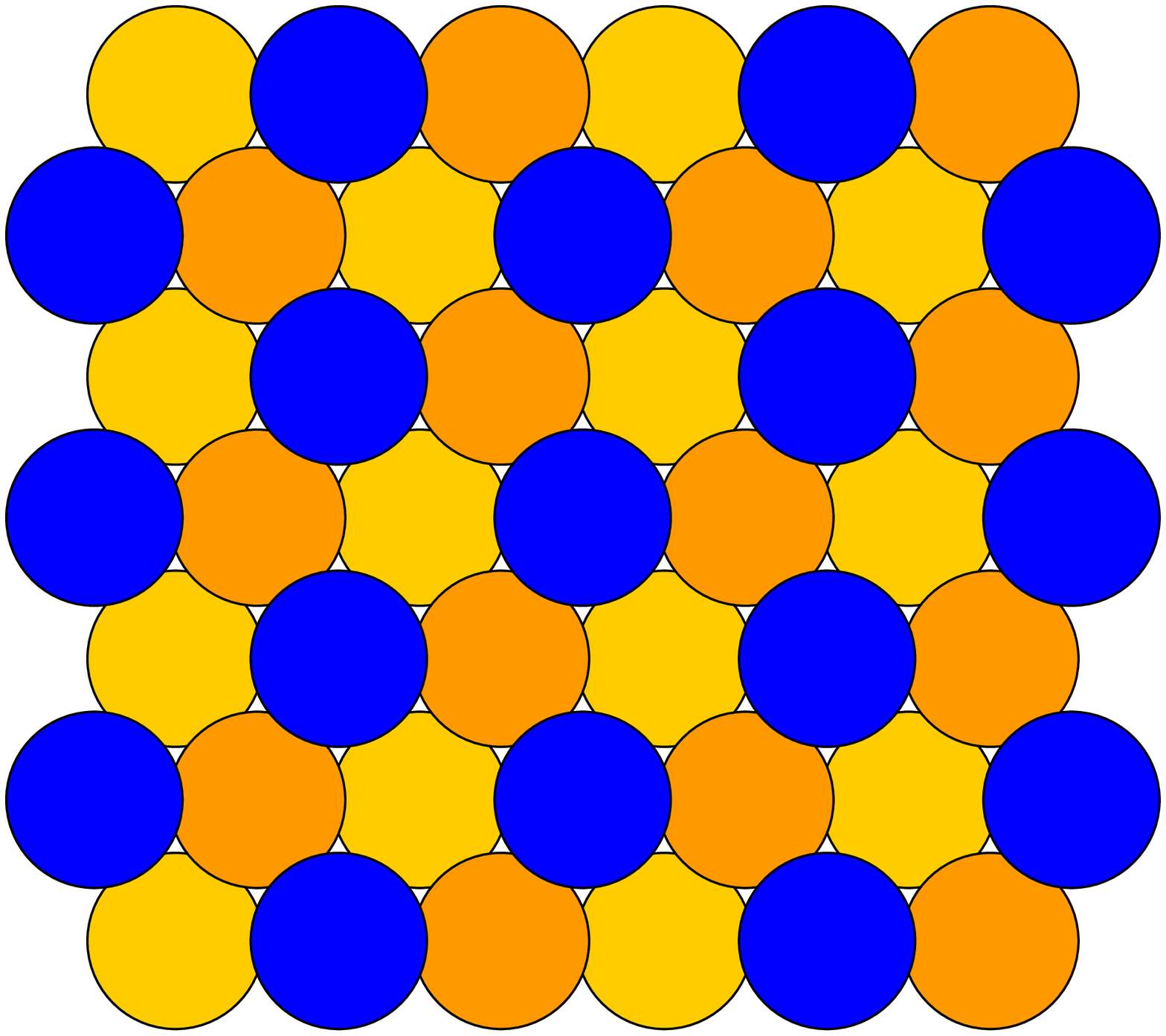}%
\caption{(Color online) Structure diagrams for an adatom and a 
  monolayer on an Au(111) surface. The 3$d$ atoms are
  represented by blue (dark) circles, various shades of orange (grey)
  represent Au atoms in different layers.}
\label{fig-geom}
\end{figure}

We investigated Fe, Co, and Ni adatoms and monolayers on the fcc
Au(111) surface.  The corresponding structures are shown in
Fig.~\ref{fig-geom}.  To get proper interatomic distances, we
performed geometry optimization using the {\sc vasp}
code.\cite{KH93,vasp-code} These calculations were done for slabs of
three layers of substrate atoms covered either by a complete layer of
3$d$ atoms (for monolayers) or by a 3$\times$3 surface supercell of
3$d$ atoms (for adatoms).  The positions of the substrate atoms in the
two lowermost layers were fixed while the positions of topmost
substrate atoms and 3$d$ atoms were relaxed.  This led to a mild
buckling of the topmost Au layer for the adatoms (about 0.02~\AA),
which we ignored in subsequent KKR-Green function calculations.  
\new{ Using a three layers thick slab instead of a semi-infinite
  substrate is justified if one is interested in relaxing the
  positions of the 3$d$ atoms above the host. However, using it for
  evaluating the MAE would be inappropriate --- for that, either a
  much thicker slab (as in Ref.~\onlinecite{SBME10}) or a proper
  semi-infinite crystal (as in this work) should be employed. }

The
optimized structural parameters as we took them from {\sc vasp}
calculations and used in the {\sc sprkkr} calculations are summarized
in Tab.~\ref{tab-geom}.  As concerns the distances between Au
substrate layers, we used the bulk interatomic distance
2.396~\AA\ everywhere except for the distance for the topmost Au
layer, which we took 2.431~\AA\ for adatoms and 2.427~\AA\ for
monolayers (as obtained via the {\sc vasp} calculations).

\begin{table}
\caption{Vertical distances $z_{\text{3d-Au}}$ in \AA\ between the
  layer containing the 3$d$ atoms and the nearest layer containing Au
  atoms.}
\label{tab-geom}
\begin{ruledtabular}
\begin{tabular}{ldd}
     &  
  \multicolumn{1}{c}{$z_{\text{3d-Au}}$} & 
  \multicolumn{1}{c}{$z_{\text{3d-Au}}$} \\
  \multicolumn{1}{c}{3$d$}   &  
  \multicolumn{1}{c}{adatom} & 
  \multicolumn{1}{c}{monolayer} \\
\hline 
Fe  &  1.889  &  2.088 \\
Co  &  1.856  &  2.035 \\
Ni  &  1.820  &  2.016 \\
\end{tabular}
\end{ruledtabular}
\end{table}


\section{Results}   \label{sec-res}


\subsection{MAE and magnetic moments}    \label{sec-moms}

\begin{table}
\caption{Magnetic properties of 3$d$ adatoms and monolayers on
  Au(111). The first two columns identify the system, the third column
  shows the MAE obtained as a difference of total energies (in meV per
  3$d$ atom), the fourth column shows spin magnetic moments for
  \mm{\bm{M}\| \bm{\hat{z}}}, and the fifth and sixth columns show
  orbital magnetic moments for \mm{\bm{M}\| \bm{\hat{z}}} and
  \mm{\bm{M}\| \bm{\hat{x}}}, respectivelly.  Magnetic moments are in
  units of \mB.}
\label{tab-mae}
\begin{ruledtabular}
\begin{tabular}{lldddd}
   &  & 
    \multicolumn{1}{c}{$E_{\text{MCA}}$} & 
    \multicolumn{1}{c}{$\mu_{\text{spin}}^{(z)}$} & 
    \multicolumn{1}{c}{$\mu_{\text{orb}}^{(z)}$} & 
    \multicolumn{1}{c}{$\mu_{\text{orb}}^{(x)}$}  \\
\hline 
Fe & adatom    &  4.07  &  3.40  &  0.536  &  0.062  \\
   & monolayer &  0.97  &  3.08  &  0.127  &  0.073  \\  [0.5ex]
Co & adatom    &  4.42  &  2.13  &  0.218  &  0.206  \\
   & monolayer &  0.42  &  2.01  &  0.156  &  0.168  \\  [0.5ex]
Ni & adatom    & -1.63  &  0.67  &  0.063  &  0.158  \\
   & monolayer & -1.97  &  0.73  &  0.118  &  0.191  \\
\end{tabular}
\end{ruledtabular}
\end{table}

The results obtained for the MAE and magnetic moments are presented in
Tab.~\ref{tab-mae}.  One can see that the easy axis of the
magnetization is perpendicular to the surface for Fe and Co adatoms
and monolayers and parallel to the surface for Ni adatom and
monolayer.  The magnetic moments were evaluated within atomic spheres
around the 3$d$~atoms.  Magnetic moments for Au atoms are small.  In
case of adatoms, the total magnetic moment induced in the Au substrate
amounts to about 5~\% of the 3$d$~adatom moment and is oriented
parallel to the moment of the adatom.  In the case of monolayers, the
total magnetic moment induced in the Au substrate per a 3$d$~atom is
about 2~\% of the 3$d$~atom moment and is oriented antiparallel to the
moments of the 3$d$ atoms.

The spin moments \ms\ practically do not depend on the magnetization
direction, while the orbital moments \mo\ strongly depend on it.  For
Fe and Ni atoms, \mo\ is significantly larger if $\bm{M}$ is parallel
to the easy axis of the magnetization than if $\bm{M}$ is parallel to
the hard axis --- in agreement with the expectations based on 
\sopt.\cite{Bru89,vdL98,ABS+06} Surprisingly, this is not the case for
Co, where for the adatom the value of \mo\ only slightly depends on
the $\bm{M}$ direction and for the monolayer the trend is even
reversed.


\subsection{Density of states}    \label{sec-dos}

\begin{figure*}
\includegraphics[width=12.9cm,viewport=0.2cm 0.3cm 17.8cm 16.0cm]{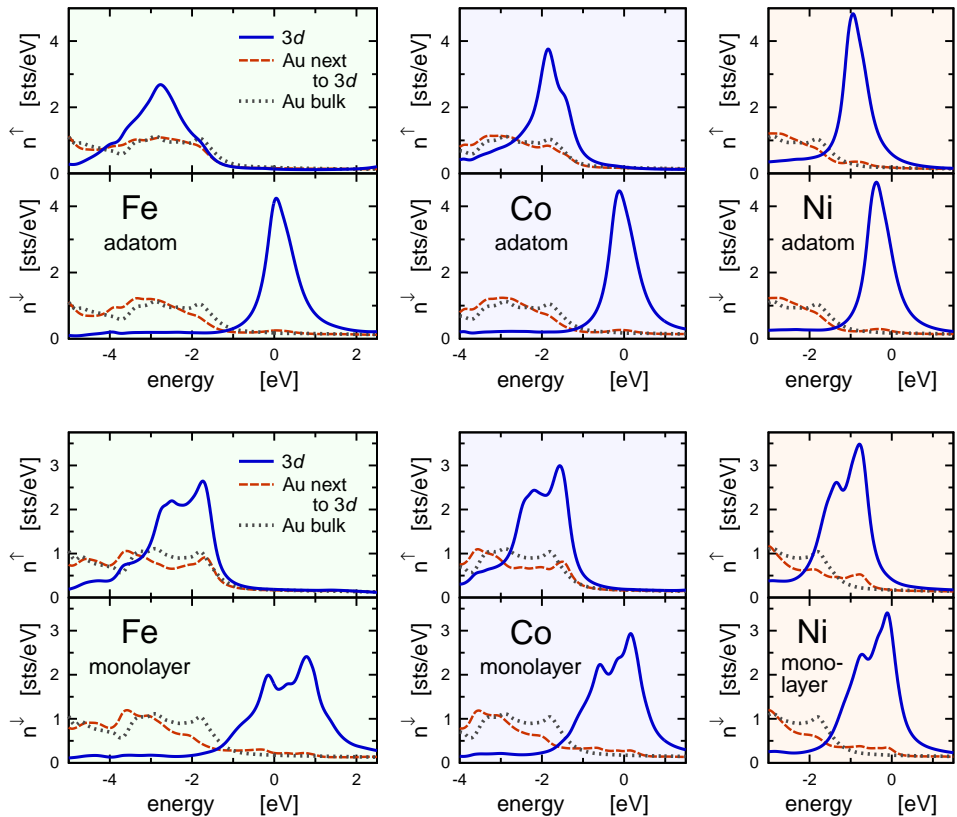}%
\caption{(Color online) Spin-projected DOS for 3$d$ adatoms and
  monolayers on Au(111) (in states per eV) for
  \mm{\bm{M}\|\bm{\hat{z}}}.  Full lines represent the DOS related to 3$d$
  atoms, dashed lines represent the DOS related to those Au atoms which
  are nearest neighbors to 3$d$ atoms, dotted lines represent the DOS for
  bulk Au.}
\label{fig-dos}
\end{figure*}

We first look at the spin-projected density of states in a range
covering the whole valence region.  This is presented in
Fig.~\ref{fig-dos}.  The data correspond to \mm{\bm{M}\|
  \bm{\hat{z}}}\ but the plot would look practically the same also for
\mm{\bm{M}\| \bm{\hat{x}}}\ at this scale.  There is a considerable
overlap between 3$d$ majority-spin states and Au states, implying that
majority-spin states are affected by hybridization while minority-spin
states are more atomic-like.

One can see that the majority-spin states are nearly fully occupied.  The
Fermi level $E_{F}$ is around the middle of the minority-spin band.
Thus if we are interested in possible effects of shifting the states
across $E_{F}$, we should focus on the influence of the SOC on
the minority-spin states. Restricting ourselves to the minority-spin
states will greatly simplify further analysis without missing the
important aspects. 
 
\begin{figure}
\includegraphics[width=8.6cm,viewport=0.2cm 0.3cm 13.0cm 20.0cm]{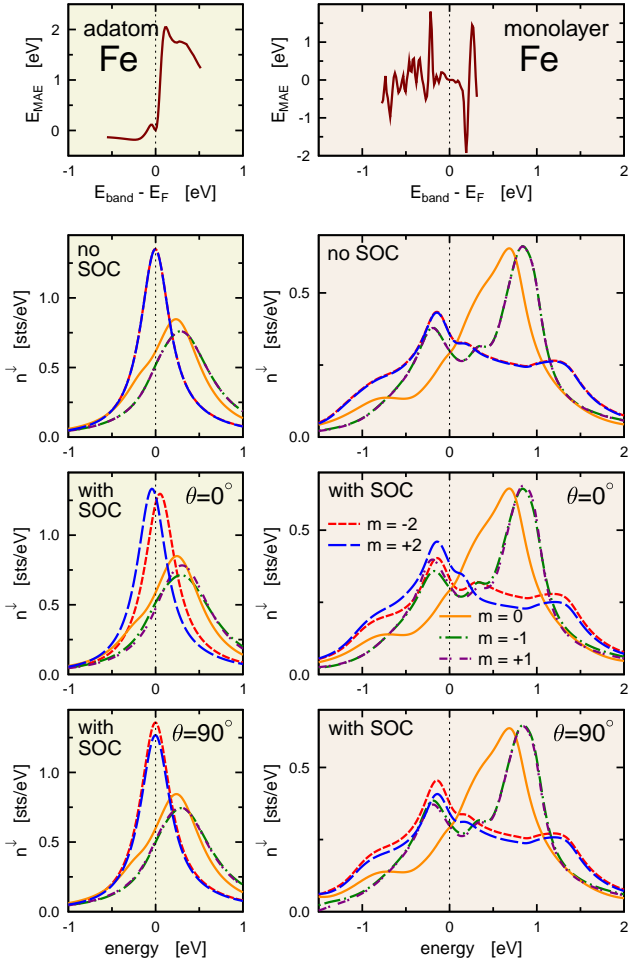}%
\caption{ (Color online) The $d$-component of the minority-spin DOS
  for a Fe adatom (left) and a Fe monolayer (right) on Au(111),
  resolved according to the magnetic quantum number $m$.  The case
  when the SOC is suppressed is presented together with the case when
  the SOC is included. The magnetization is either perpendicular to
  the surface ($\theta$=0$^{\circ}$) or parallel to the surface
  ($\theta$=90$^{\circ}$).  The dependence of the MAE on the position
  of the top of the valence band is shown in the top graphs. }
\label{fig-glob-Fe}
\end{figure}

Studying how $m$-resolved DOS varies upon the rotation of the
magnetization requires some clarifications.  The projection of the DOS
according to the quantum number $m$ has to be done always in the same
reference frame, disregarding the orientation of \MM.  We call this
frame the ``global reference frame'' --- it is fixed to the underlying
crystal lattice.  If the $m$-projections are done in different
reference frames for different magnetization directions, the
definitions of the $m$-components themselves also vary, because they
are linked to the spherical harmonics $Y_{\ell m}$ which are defined
with respect to the $x$, $y$, $z$ axes.  On the other hand, if one
wants to retain and emphasize the difference between spin-up and
spin-down contributions to the DOS, one has to make the projection in
a ``local reference frame'', rotated so that the $z$~axis coincides
with the magnetization direction.  The need for this can be easily
seen from the Stern-Gerlach term in the Pauli equation, $\bm{\sigma}
\cdot \bm{B}$, which is diagonal only if the effective magnetic field
$\bm{B}$ is parallel to the $z$ axis.  If the spin quantization axis
is not parallel to the magnetization direction, the chosen
representation strongly mixes spin-up and spin-down components.

These two circumstances suggest that if one wants to study the DOS for
different directions of the magnetization \MM, one has to renounce
either having a universal definition of the $m$-projections or
retaining well-separated spin-resolved DOS components.  This is not an
issue if the SOC is ignored because then the direction of the
magnetization has no effect on the electronic structure anyway.
However, if the SOC is accounted for and the dependence of the DOS on
the direction of \MM\ is in focus, this is a serious obstacle.

\begin{figure}
\includegraphics[width=8.6cm,viewport=0.2cm 0.3cm 13.0cm 20.0cm]{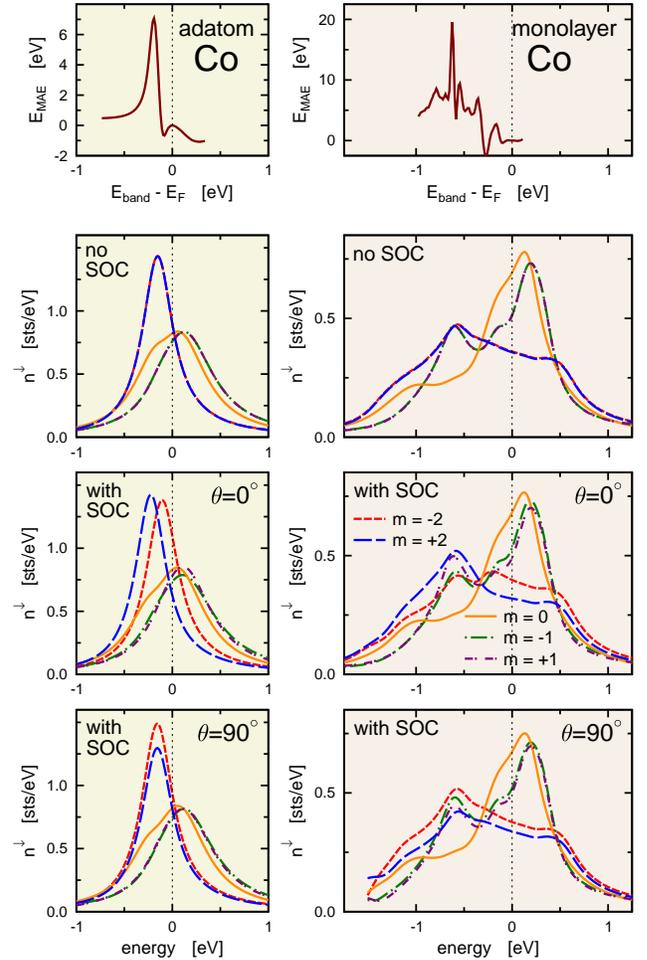}%
\caption{ (Color online) Density of the minority-spin $d$ states resolved
  according to the magnetic quantum number $m$ and the dependence of
  the MAE on the position of the top of the valence band for a Co
  adatom (left graphs) and a Co monolayer (right graphs) on
  Au(111). Otherwise, this figure is analogous to
  Fig.~\protect\ref{fig-glob-Fe}. } 
\label{fig-glob-Co}
\end{figure}

\begin{figure}
\includegraphics[width=8.6cm,viewport=0.2cm 0.3cm 13.0cm 20.0cm]{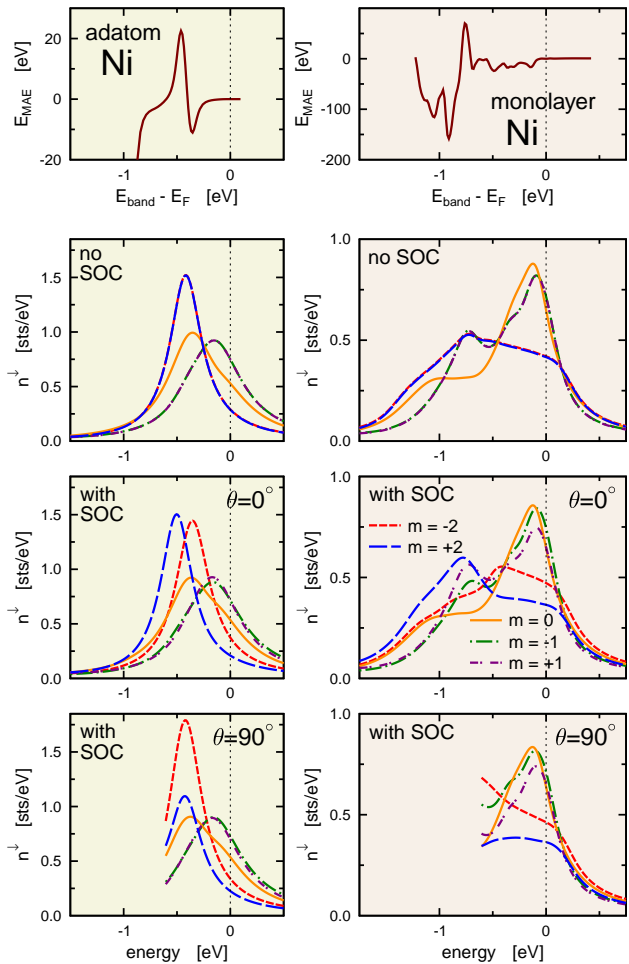}%
\caption{ (Color online) Density of the minority-spin $d$ states
  resolved according to the magnetic quantum number $m$ and the
  dependence of the MAE on the position of the top of the valence band
  for a Ni adatom (left graphs) and a Ni monolayer (right graphs) on
  Au(111). Otherwise, this figure is analogous to
  Fig.~\protect\ref{fig-glob-Fe}. }
\label{fig-glob-Ni}
\end{figure}

Fortunately, this restriction can be by-passed in our case.  It is
possible to get well-defined spin-minority DOS $m$-decomposed in a
global reference frame even if \MM\ is is not parallel to the $z$
axis, relying on an approximate procedure which is described in
appendix~\ref{sec-proj}.  The procedure combines results for an
$m$-decomposition in global and local reference frames.  Employing
this technique, we obtained the density of minority-spin $d$-states
resolved according to the magnetic quantum number $m$ as shown in
Figs.~\ref{fig-glob-Fe}--\ref{fig-glob-Ni}.  The magnetization is
either out-of-plane ($\bm{M}\|\bm{\hat{z}}$, $\theta$=0$^{\circ}$) or
in-plane ($\bm{M}\|\bm{\hat{x}}$, $\theta$=90$^{\circ}$) and the
$m$-projections are defined in the same (global) reference frame in
both cases.  To highlight the effect of the SOC, we present results
obtained with SOC suppressed and with SOC accounted for.

It can be seen readily from the plots in
Figs.~\ref{fig-glob-Fe}--\ref{fig-glob-Ni} that if the SOC is
suppressed, the DOS does not depend on the sign of $m$.  Components
for $+|m|$ and $-|m|$ are the same in this case, the only splitting
comes from the crystal field.  If the SOC is taken into account, then
the DOS further depends on whether $m$ is positive or negative.  There
is a significant difference in how the $\pm|m|$ states are split for
out-of-plane magnetization and for in-plane magnetization (especially
for the $m$=$\pm$2 case).

The procedure outlined in appendix~\ref{sec-proj} can be applied only
if the SOC-induced splitting of the majority-spin states is negligible
in the energy region in which we are interested in, i.e., around
\EF. This assumption is well justified for Fe and Co.  However, it is
not so good for Ni, where the exchange-splitting is quite small
(cf.~Fig.~\ref{fig-dos}) and the influence of the SOC on the
majority-spin DOS is significant up to about 0.5~eV below $E_{F}$.
Therefore, for Ni we present the data only for for \mm{E>-0.6}~eV and
even there they are less reliable than analogous data in
Figs.~\ref{fig-glob-Fe}--\ref{fig-glob-Co}.  The full energy range for
$\bm{M}\|\bm{\hat{x}}$ is covered by
Figs.~\ref{fig-loc90-Fe}--\ref{fig-loc90-Ni} in
appendix~\ref{sec-locbas}, where we present the $m$-resolved DOS
projected in a local reference frame rotated so that the
$z^{\text{(loc)}}$~axis is parallel to $\bm{M}$.  
\new{ (For $\bm{M}\|\bm{\hat{z}}$, the global reference frame and the
  local reference frame coincide, because $z^{\text{(loc)}}$ is then
  identical with $z$.)  }

\new{ The definitions of individual $m$-components employed in
  appendix~\ref{sec-locbas} and employed in this section obviously
  differ. }
  One cannot, therefore, directly compare
the plots where the DOS was resolved into $m$-components in the global
reference frame (Figs.~\ref{fig-glob-Fe}--\ref{fig-glob-Ni}) with
plots where the DOS was resolved in the local reference frame
(appendix~\ref{sec-locbas}).  What is common in both reference frames
is that the SOC-induced splitting of the $\pm|m|$ components it
significantly smaller for $\bm{M}\|\bm{\hat{x}}$\ than for
$\bm{M}\|\bm{\hat{z}}$.

Let us summarize the picture obtained by inspecting the DOS.  First,
note that the minority-spin DOS for the adatoms has quite an atomic
character: if the SOC is suppressed, it can be seen as representing
just three broadened atomic levels, depending on $|m|$.  For
monolayers, the hybridization between states from different 3$d$ atoms
is considerable, so the DOS does not have an atomic character any more.
The second point to emphasize is that the influence of the SOC is
significantly larger for $\theta$=0$^{\circ}$ than for
$\theta$=90$^{\circ}$.  More specifically, for $\theta$=0$^{\circ}$
the SOC splits the $m$=$\pm$2 peak into two and shifts their positions
in different directions while for $\theta$=90$^{\circ}$, the peak
positions remain the same (only their intensities change).

The splitting of $m$-resolved DOS peaks by the SOC suggests that the
MAE could be very sensitive to the position of $E_{F}$ with respect to
these peaks.  Therefore we calculated the MAE while varying the
position of the top of the valence band $E_{\text{band}}$, i.e., the
band filling.  The results are shown in top panels of
Figs.~\ref{fig-glob-Fe}--\ref{fig-glob-Ni}.  One can see that for the
adatoms there is indeed a sharp peak in the MAE just at the energy
where there is a peak for the $|m|$=2 component in the case of no SOC.
This is especially visible for Co and Ni adatoms.  For the Fe adatom
this aspect is overshadowed by another strong feature stemming from
the fact that, in this case, also the $|m|$=1 states are affected by
SOC.  The situation for monolayers is more complicated because the
$m$-components are not atomic-like any more.  Nevertheless, even here
a strong peak in the curve for $E_{\text{MAE}}$ as a function of
$E_{\text{band}}$ appears at the energy where the DOS components for
$|m|$=2 have their maximum.  We would like to note in this context
that the density of the $E_{\text{band}}$ mesh used in the calculation
is the same for adatoms and monolayers.  This means that the
observation that the $E_{\text{MAE}}(E_{\text{band}})$ oscillations
are much wilder for monolayers than for adatoms describes a real
effect. Probably this is connected with hybridization between 3$d$
atoms which is present for monolayers but absent for adatoms.


\subsection{Effect of SOC on the energy levels via model Hamiltonian}     
\label{sec-model}

We could see in Sec.~\ref{sec-dos} how the SOC splits electronic states
for different orientations of the magnetization.  Let us check to what
extent this can be described within a simple model with only the
crystal-field effects taken into account. This corresponds to a
situation where the electron feels only the Coulombic field generated
by charges located at the positions of the nuclei.

\begin{table}
\caption{Energy levels of $d$ electrons for an axial-crystal-field
  Hamiltonian if there is no exchange splitting and no SOC.}
\label{tab-levels}
\begin{ruledtabular}
\begin{tabular}{lll}
energy   & \hspace*{2mm} & $Y_{\ell m}$   \\
\hline
$E_{1} = P$        & & $m$=$\pm$2     \\
$E_{2} = {}-2(P+Q)$  & & $m$=0        \\
$E_{3} = Q$        & & $m$=$\pm$1     \\
\end{tabular} 
\end{ruledtabular}
\end{table}

To highlight the essential features, we restrict ourselves to $d$
electrons in an axial field (corresponding to $D_{4d}$, i.e.,
antiprism symmetry).  If there is no magnetic order or SOC, the
crystal-field Hamiltonian is given as described for example in the
book of Bersuker\cite{Ber10} [Eqs.~(4.9)--(4.10) and Tab.~4.1].  The
Hamiltonian is determined by two parameters (if the constant energy
shift is omitted), resulting in three spin-degenerate energy levels as
given in Tab.~\ref{tab-levels}.  The order of levels $E_{1}$, $E_{2}$,
and $E_{3}$ depends on the values of parameters $P$ and $Q$.  Levels
$E_{1}$ and $E_{3}$ are double degenerate.  Non-zero terms of the
crystal-field Hamiltonian are
\begin{equation}
H^{(\text{cry})}_{ms,m's'} \, = \, \left\{
\begin{array} {c@{\qquad}l@{\quad}r}
P       &  m=\pm2,  &  s=s'  \;\; ,\\
Q       &  m=\pm1,  &  s=s'  \;\; ,\\
-2P-2Q  &  m=0,     &  s=s'  \;\; .
\end{array}
\right.
\end{equation}
The subscript $ms$ combines the magnetic quantum number $m$ and the
spin quantum number $s$, meaning that our Hamiltonian is represented
by a 10$\times$10 matrix.

The magnetization of the system is reflected by the exchange field
Hamiltonian $H^{(\text{ex})}$.  To distinguish between two
orientations of the magnetization, we keep the spin quantization axis
fixed (parallel to $z$) and vary the Hamiltonian $H^{(\text{ex})}$.
The non-zero terms of $H^{(\text{ex})}$ for \mm{\bm{M}\| \bm{\hat{x}}}
are
\begin{equation}
H^{(\text{ex})}_{ms,m's'} \, = \, B \qquad m=m', \quad s\neq s'
\end{equation}
and for \mm{\bm{M}\| \bm{\hat{z}}} they are 
\begin{subequations}
\label{eq-hex}
\begin{eqnarray}
H^{(\text{ex})}_{ms,m's'} \, = \, B  & \quad m=m', & \, s=s'=-1/2 \;\; ,\\
H^{(\text{ex})}_{ms,m's'} \, = \, {-B} & \quad m=m', & \, s=s'=+1/2 \;\; .
\end{eqnarray}
\end{subequations}

The third contribution to the model Hamiltonian comes from the SOC.
The spin quantization axis is kept parallel to $z$, so the 
Hamiltonian \mm{H^{(\text{SOC})}=\xi\bm{L}\cdot\bm{S}}\ can be
represented as (cf.\ St\"{o}hr)\cite{StSi+06}
\begin{equation}
H^{(\text{SOC})} \, = \, 
 \left[ \begin {array}{cccccccccc} {-\xi}&0&0&0&0&0&\xi&0&0&0
\\\noalign{\medskip}0&{-\frac{\xi}{2}}&0&0&0&0&0&\frac{\xi\sqrt{6}}{2}&0&0
\\\noalign{\medskip}0&0&0&0&0&0&0&0&\frac{\xi\sqrt{6}}{2}&0
\\\noalign{\medskip}0&0&0&\frac{\xi}{2}&0&0&0&0&0&\xi\\\noalign{\medskip}0&0
&0&0&\xi&0&0&0&0&0\\\noalign{\medskip}0&0&0&0&0&\xi&0&0&0&0
\\\noalign{\medskip}\xi&0&0&0&0&0&\frac{\xi}{2}&0&0&0\\\noalign{\medskip}0&
\frac{\xi\sqrt{6}}{2}&0&0&0&0&0&0&0&0\\
\noalign{\medskip}0&0&\frac{\xi\sqrt{6}}{2} 
&0&0&0&0&0&{-\frac{\xi}{2}}&0\\\noalign{\medskip}0&0&0&\xi&0&0&0&0&0
&{-\xi}\end {array} \right] 
\;\; .
\end{equation}
The total Hamiltonian we have to diagonalize is 
\begin{equation}
H \, = \, H^{(\text{cry})} \, + \, H^{(\text{ex})} \, + \, H^{(\text{SOC})}
\;\; .
\label{modham}
\end{equation}

\begin{table}
\caption{Parameters (in eV) for the model Hamiltonian simulating 3$d$
  adatoms on Au(111) by means of an axial crystal field model.}
\label{tab-params}
\begin{ruledtabular}
\begin{tabular}{cccc}
 & Fe & Co & Ni \\
\hline
$\Delta_{1}$               & 0.24  & 0.21   & 0.06   \\
$\Delta_{2}$               & 0.05  & 0.06   & 0.20   \\
$E_{\downarrow} - E_{\uparrow}$ & 2.81  & 1.96   & 0.57   \\
$\xi$                     & 0.065  & 0.085  & 0.108  \\
\end{tabular} 
\end{ruledtabular}
\end{table}
 
\begin{figure*}
\includegraphics[viewport=0.2cm 0.2cm 17.7cm 8.7cm]{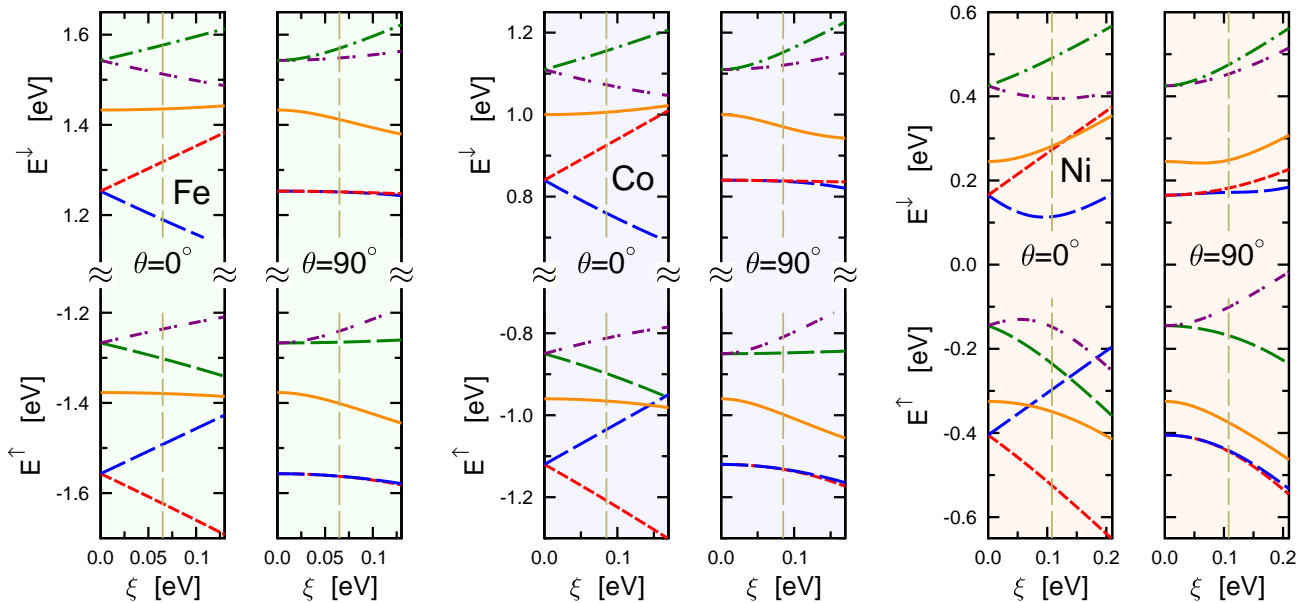}%
\caption{(Color online) Dependence of eigen-energies of the model
  Hamiltonian given by Eq.~(\ref{modham}) on the SOC strength $\xi$
  for two orientations of $\bm{M}$.  The model parameters for Fe, Co,
  and Ni adatoms are given in Tab.~\protect\ref{tab-params}. Thin
  dashed lines mark $\xi$ values appropriate for each element. The
  zero of energy corresponds to the case with no magnetization and no
  crystal field.}
\label{fig-model}
\end{figure*}

We want to apply this model for an adatom, where the hybridization is
small and the crystal-field effects will be important.  Looking at the
adatom-related panels of Figs.~\ref{fig-glob-Fe}--\ref{fig-glob-Ni},
we can see that in the absence of SOC the minority-spin DOS indeed
resembles three atomic-like energy levels, as in
Tab.~\ref{tab-levels}.  It is convenient to introduce level spacings
\begin{eqnarray}
\Delta_{1} & \equiv & E_{2} - E_{1}  \;\; , \\
\Delta_{2} & \equiv & E_{3} - E_{2}  \;\; ,
\end{eqnarray}
through which we can express the model Hamiltonian parameters as 
\begin{eqnarray}
P & = & {}-\frac{3 \Delta_{1} + 2 \Delta_{2}}{5} \;\; , \\
Q & = & \frac{2 \Delta_{1} + 3 \Delta_{2}}{5} \;\; . 
\end{eqnarray}
To simulate 3$d$ adatoms on Au(111), one should read the $\Delta_{1}$,
$\Delta_{2}$ splittings from
Figs.~\ref{fig-glob-Fe}--\ref{fig-glob-Ni} to get the values for the
parameters $P$, $Q$ and the exchange splitting from Fig.~\ref{fig-dos}
to get the parameter $B$ using $E_{\downarrow}-E_{\uparrow}=2B$.  The
SOC parameters $\xi$ can be obtained via {\em ab-initio}
calculations.\cite{DWW+88} The appropriate values are given in
Tab.~\ref{tab-params}.

A general idea how the SOC affects the energy levels can be obtained
by diagonalizing the Hamiltonian~(\ref{modham}) for different
orientations of $\bm{M}$ while the SOC strength $\xi$ is gradually
increased from zero to a realistic value.  The corresponding results
are presented in Fig.~\ref{fig-model} where we show energy levels
$E^{\uparrow}(\xi)$ and $E^{\downarrow}(\xi)$ for parameters given in
Tab.~\ref{tab-params}.  The ``proper'' value of $\xi$ for each element
is marked by a thin dashed line.  To avoid confusion, we should note
that our categorizing of levels as $E^{\uparrow}$ or $E^{\downarrow}$
is done just for convenience, by comparing their positions to the
spin-projected DOS shown in Fig.~\ref{fig-dos}.  We care only about
the energy levels in this context and not about the wave functions, so
the issue of mixed spin character for $\theta$=90$^{\circ}$, discussed
in Sec.~\ref{sec-dos} and in appendix~\ref{sec-proj}, does not
interfere with our analysis.

A prominent feature of Fig.~\ref{fig-model} is that the effect of
$\xi$ is much less for in-plane magnetization ($\theta$=$90^{\circ}$)
than for perpendicular magnetization ($\theta$=$0^{\circ}$).  This is
especially true for the lowest energy which corresponds to $m$=$\pm$2.
By comparing this observation to
Figs.~\ref{fig-glob-Fe}--\ref{fig-glob-Ni}, we see that the simple
crystal-field model indeed accounts for the trends in the $m$-resolved
DOS for the 3$d$ adatoms.  It is worth noting that if the
exchange-field parameter $B$ decreases (i.e., going from Fe to Co to
Ni), the $m$=$\pm$2 energy levels split also for the
$\theta$=$90^{\circ}$ case (in-plane magnetization).  A similar trend
can be seen also in the DOS in Sec.~\ref{sec-dos}: the difference
between $m$=$\pm$2 curves in the lowermost left panels in
Figs.~\ref{fig-glob-Fe}--\ref{fig-glob-Ni} increases when going from
Fe to Co to Ni.


\section{Discussion}   \label{sec-discuss}

Our aim was to investigate whether markers of MAE can be seen in
intuitive quantities such as the $m$-resolved DOS.
Figs.~\ref{fig-glob-Fe}--\ref{fig-glob-Ni} (in conjunction
with Figs.~\ref{fig-loc90-Fe}--\ref{fig-loc90-Ni}) show how SOC
affects the DOS depending on the orientation of the 
magnetization~$\bm{M}$.  The corresponding changes in the DOS can be
linked to the \mca\ of adatoms.  Particularly for the Fe and Co adatoms
one can see that for $\theta$=0$^{\circ}$ the SOC splits the
$|m|$=2~component of the DOS in such a way that one of the peaks is
pushed above $E_{F}$ (or at least an essential part of it).  The
band-energy contribution to the total energy is thus substantially
reduced.  As this effect does not occur for $\theta$=90$^{\circ}$, the
out-of-plane orientation of \MM\ is energetically more favored and
the corresponding MAE is positive, in agreement with Tab.~\ref{tab-mae}

The SOC-induced splitting of the $|m|$=2 peak occurs for
$\theta$=0$^{\circ}$ also for the Ni adatom.  However, in that case
both peaks remain below \EF\ and the change in the band-energy is
therefore much smaller.  The influence of SOC for the
$\theta$=90$^{\circ}$ case is best seen if the $m$-projection is done
in a rotated local reference frame, as in Fig.~\ref{fig-loc90-Ni}.
This is because the isolation of the minority-spin DOS in the global
reference frame cannot be done properly due to the small energy
separation between the majority-spin and minority-spin states of Ni.
The lowermost graphs in Fig.~\ref{fig-glob-Ni} have to be seen as
primarily illustrative in this respect because they are affected by
the fact that majority-spin states are still influenced by the SOC in
this region.  Focusing on the unambiguous data in
Fig.~\ref{fig-loc90-Ni} one can see that for $\theta$=90$^{\circ}$ the
states with $|m^{\text{(loc)}}|$=2 are split in such a way that part
of one of the SOC-split peaks is pushed above \EF.  This effect
overruns the corresponding effect on the $|m|$=2 states for
$\theta$=0$^{\circ}$ (Fig.~\ref{fig-glob-Ni}) and, accordingly, the
easy axis of magnetization is in-plane for the Ni adatom.

Effects of this kind can hardly be identified for monolayers.  In this
case the hybridization between the 3$d$ states distorts the
atomic-like character of the states and one would have to consider a
lot of contributions, similarly as if the $E(\bm{k})$ band-structure
of layered systems is analyzed.\cite{WWF93a,DKS+94,GC+12}

The simple crystal field model accounts qualitatively for many aspects
of the \mca\ of adatoms, indicating that this anisotropy can be
understood intuitively as an interplay between the axial crystal
field, the exchange field and the spin-orbit coupling.  However, there
are also differences between the pictures offered by the model
Hamiltonian and by the DOS obtained from {\em ab initio} calculations.
For example, the model Hamiltonian suggests that for an in-plane
magnetization ($\theta$=$90^{\circ}$), the splitting between the
$m$=$\pm$1 levels is larger than the splitting between the $m$=$\pm$2
levels (Fig.~\ref{fig-model}); however, we do not observe this feature
in Figs.~\ref{fig-glob-Fe}--\ref{fig-glob-Ni}.  This means that
effects not included in the simple model of Sec.~\ref{sec-model}, such
as hybridization, are important as well.

It should be noted that by monitoring SOC-induced changes in the DOS
one accounts only for the band-energy contribution to the total
energy, omitting thus the terms that explicitly depend on the change
of the potential upon rotation of \MM\ (see, e.g., chapter 6 of the
monograph of Weinberger\cite{Wei08} for more details).  This is
equivalent to relying on the so called force theorem.  If the MAE is
evaluated accounting for the band energy contribution only (by means
of the torque method),\cite{WWW+96,SSB+06} we obtain
$E_{\text{MCA}}$=5.7~meV for the Fe adatom, 1.9~meV for the Co adatom,
and \mm{{-0.8}}~meV for the Ni adatom.  Comparison with
Tab.~\ref{tab-mae} that gives $E_{\text{MCA}}$ as a difference of
total energies shows that the change in the band energy does not fully
account for the \mca\ but nevertheless constitutes a significant part
of it.  One should also keep in mind that the SOC-induced splitting of
the DOS is not the only way the band energy is changed upon rotation
of \MM.  For example, all effects contained in Eq.~(\ref{2nd})
contribute as well.  Accordingly, what has been done here is
identifying and visualizing one important mechanism contributing to
the \mca.  We suggest (following earlier
hints)\cite{WWF93b,DKS+94,MHB+96,RKF+01,KS+15} that this mechanism may
be the dominant one for some adatoms and small clusters on surfaces
--- including those that attracted a lot of attention
recently.\cite{DDA+13,DSS+14,RBR+14,BDS+15}

Another interesting system to be mentioned in this context is lithium
nitridoferrate Li$_{2}$[(Li$_{1-x}$Fe$_{x}$)N] which attracted a lot
of attention due to its very high \mca.\cite{JMT+13,NW+02,AA+14,KS+15}
This system can be viewed as a collection of semi-isolated Fe
impurities.  A similar effect as the one investigated here could
therefore be important for Li$_{2}$[(Li$_{1-x}$Fe$_{x}$)N] and
attention was indeed paid to it in this respect.\cite{AA+14,KS+15}
Generally, the mechanism we explored here should be important whenever
the width of the electronic bands becomes comparable to the SOC-induced
changes in the orbital-resolved DOS upon the rotation of the
magnetization.

If the \mca\ is generated via pushing some SOC-split levels across the
Fermi level, it must crucially depend on their mutual
positions. Specifically in our case, it must depend on the position of
the energy bands of the adatom with respect to the Fermi level of the
substrate (cf.~also the top graphs of
Figs.~\ref{fig-glob-Fe}--\ref{fig-glob-Ni}).  Therefore, one might be
able to manipulate the MAE by changing the substrate $E_{F}$, e.g.,
via doping.

Even though the aim of this study is not to reproduce experimental MAE
for specific systems, it is useful to compare our values of MAE with
available experiments.  There are no data for adatoms on Au(111) but
there have been several experimental studies of Fe and Co layers on
Au(111).  Before comparison with experiment is done, the dipole or
shape anisotropy energy for monolayers must be given.  It is
${-0.18}$~meV, ${-0.08}$~meV, and ${-0.01}$~meV for Fe, Co, and Ni monolayer,
respectively.  These values are smaller than the \mca\ energy given in
Tab.~\ref{tab-mae}.  So we predict that Fe and Co monolayers on
Au(111) have out-of-plane easy axes of magnetization and a Ni
monolayer (for which there are no experiments available) has an
in-plane easy axis of magnetization.  Earlier calculations for a Co
monolayer on Au(111) predicted an in-plane easy axis of magnetization
for this system;\cite{USBW96,SBE+14} the reason for the difference is
almost certainly the structural relaxation which was accounted for
here but not in the two earlier works.

Despite several experimental studies of \mca\ of Fe and Co layers on
Au(111) done in the past, drawing conclusions from them is not easy or
unambiguous because the growth conditions vary and typically do not
favor formation of a single monolayer.  A critical analysis of
experiments is beyond our scope.  For a Fe monolayer it is probably 
safe to say that experiment suggests an out-of-plane easy
axis,\cite{LRP+93,TPA+04,LCH+09,AD+10} as our calculations do.  For a
Co monolayer, the situation is more complicated.  For bilayer islands
on Au(111) one gets an out-of-plane easy
axis.\cite{DDD+99b,RRS+07,TEM+15} Again growth conditions may be
crucial.\cite{PSC+00} No data seem to exist for a single monolayer on
Au(111).  As a whole, even though we cannot verify our results by a
comprehensive comparison with experiment, agreement with available
data as well as the fact that our values of MAE are in the same range
of values as those obtained for similar systems indicate that our
results are reliable and can be used as a basis for the analysis we
performed in Secs.~\ref{sec-dos} and~\ref{sec-model}.


\section{Conclusions}   \label{sec-zaver}

The effect of spin-orbit coupling on adatoms that only weakly
hybridize with a substrate consists in splitting  atomic-like levels
that would be degenerate in its absence.  The splitting is much larger
if the magnetization is oriented perpendicular to the surface than if
it is oriented parallel to the surface and can be viewed as a combined
result of crystal field, exchange splitting and spin-orbit coupling.
If the originally degenerate level is close to the Fermi level, one of
the peaks can be pushed above it, decreasing thereby the energy of the
system.  This effect represents a significant contribution to the
\mca\ of adatoms.  If hybridization smears out the atomic-like
character of energy levels, as it is the case for monolayers, this
effect is not so important.


\begin{acknowledgments}
This work was supported by the Grant Agency of the Czech Republic
within the project 108/11/0853, by the Deutsche Forschungsgemeinschaft
within the project SFB 689 ``Spinph\"{a}nomene in reduzierten
Dimensionen'' and by Ministry of Education, Youth and Sports within
the project CENTEM PLUS (LO1402).
\end{acknowledgments}


\appendix

\section{Spin-resolved and $m$-resolved DOS for
  \mm{\bm{M}\nparallel \bm{\hat{z}}}}
\label{sec-proj}

It was argued in Sec.~\ref{sec-dos} that if one wants to see how
individual $m$-components of the DOS are affected by the rotation of
the magnetization \MM, one should perform the $m$-projections always
in a global reference frame so that the definitions of the
$m$-components remain the same.  However, in case that
\mm{\bm{M}\nparallel \bm{\hat{z}}}, projecting the DOS in a global
reference frame mixes the spin components because the spin
quantization axis is no longer parallel to \MM.  In this appendix we
present a method to restore the separation of spin components in the
DOS even in such a case.  Our goal is achieved by a detour, combining
results of projections in the global and local reference
frames. Effectively, it could be seen as a way to make the
spin-projection and the $m$-projection in different reference frames.

Let us recall that inside an atomic sphere the DOS for a spin channel
can be represented by means of the Green function $G(E)$ as
\begin{equation}
n(E) \: = \:
  -\frac{1}{\pi} \, \Im \int \! \dstd^{3} \bm{r} \: 
\braket{ \bm{r} | G(E) | \bm{r} } 
\; .  \\
\end{equation}
We omit the spin labels here for brevity.  Angular-momentum
projections of $n(E)$ can be obtained by means of the spherical
harmonics.  These spherical harmonics $Y_{\ell m}$ can be defined in a
global reference frame (fixed to the crystal lattice) or in a local
reference frame chosen so that the $z^{\text{(loc)}}$ axis is parallel
to \MM.  The way the DOS components $n_{L}$ are defined thus depends
on the reference frame.  We can write schematically (again, for each
spin channel)
\begin{eqnarray}
n^{(\text{glo})}_{L}(E) & = & -\frac{1}{\pi} \, \Im \, \bra{Y^{(\text{glo})}_{L}} 
  G(E) \ket{Y^{(\text{glo})}_{L}}  
\;\; ,  \label{glodir} \\
n^{(\text{loc})}_{L}(E) & = & -\frac{1}{\pi} \, \Im \, \bra{Y^{(\text{loc})}_{L}} 
  G(E) \ket{Y^{(\text{loc})}_{L}}  
\;\; . \label{locdir} 
\end{eqnarray}
Integration over the radial coordinate is implicitly assumed. 

We start by projecting the DOS in the local reference frame, where
\mm{\bm{M} \| \bm{\hat{z}}^{\text{(loc)}} }.  In this way we perform
the separation of the spin components.  We assume that this separation
was performed ``once for all times'', i.e., it will be preserved
during the whole subsequent procedure.  All the manipulations will be
applied to minority-spin DOS and to majority-spin DOS separately.
 
\begin{figure}
\includegraphics[width=8.6cm,viewport=0.2cm 0.3cm 11.5cm 6.0cm]{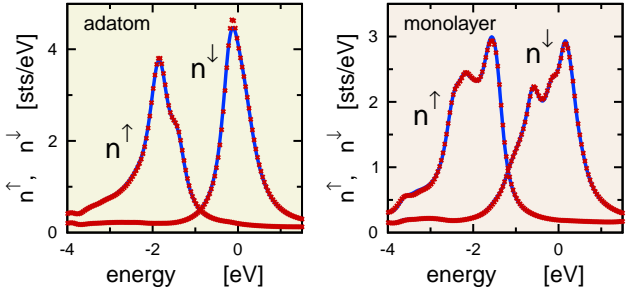}%
\caption{ (Color online) Spin-projected DOS for a Co adatom and a Co
  monolayers on Au(111) for \mm{\bm{M}\|\bm{\hat{z}}} (solid blue
  lines) and for \mm{\bm{M}\|\bm{\hat{x}}} (red cross markers). }
\label{fig-rotmag}
\end{figure}

This requires a further comment.  By doing the spin-projection in the
rotated local reference frame, we get spin-up and spin-down states
assuming that the spin quantization axis is in a general direction
while when dealing with the global reference frame, the
spin-quantization axis is fixed and parallel to $z$.  However, this
difference can be neglected in our case: we checked that the
spin-projected DOS (without any $m$-decomposion) looks practically the
same no matter whether the magnetization is in-plane or out-of-plane.
As an illustration, spin-projected DOS for a Co adatom and a Co
monolayer is shown in Fig.~\ref{fig-rotmag} for two magnetization
directions.  These spin projections were obtained in local reference
frames defined so that the $z^{\text{(loc)}}$ axis is always parallel
to \MM.  One can see that there is hardly any difference between the
DOS for \mm{\bm{M}\|\bm{\hat{x}}}\ and \mm{\bm{M}\|\bm{\hat{z}}}.
Another hint that the spin projections can be maintained upon rotating
\MM\ comes from the fact that if the magnetization is rotated, the
spin magnetic moments almost do not change.  There is a common
experience that this is the case for all magnetic systems.  By
assuming that the spin-projected DOS does not depend on the direction
of the magnetization, we make an effective decoupling of spin and
orbital degrees of freedom.  This enables us to focus on changes in
the $m$-resolved components.  A similar decoupling is used, e.g., when
deriving useful relations for the angular-dependence of the magnetic
dipole term $T_{z}$ for analyzing the x-ray magnetic circular
dichroism spectra.\cite{SK95a,Sto99}

So far we have obtained minority-spin DOS and majority-spin DOS,
$m$-resolved in the local frame.  Now we need to transform the
spin-polarized DOS from the basis spanned by $Y^{(\text{loc})}_{L}$ to
the basis spanned by $Y^{(\text{glo})}_{L}$. A straightforward
transformation between the $n^{(\text{loc})}_{L}$ and the
$n^{(\text{glo})}_{L}$ components is generally not possible --- one
always has to start with the Green function $G$ to get $n_{L}$ in a
new basis.  However, the transformation can be done provided that $G$
is diagonal in the basis in which $n_{L}$ has been initially
known. Indeed, if we assume that
\[
 \bra{Y^{(\text{glo})}_{L}} 
  G(E) \ket{Y^{(\text{glo})}_{L'}} \: = \:
\delta_{L L'} \,  \bra{Y^{(\text{glo})}_{L}} 
  G(E) \ket{Y^{(\text{glo})}_{L}}
\;\; ,
\]
we obtain 
\begin{eqnarray}
n^{(\text{loc})}_{L}(E) & = & -\frac{1}{\pi} \, \Im \, \bra{Y^{(\text{loc})}_{L}} 
  G(E) \ket{Y^{(\text{loc})}_{L}}  
\;\; ,  \nonumber  \\
 & = &  -\frac{1}{\pi} \, \Im \, \sum_{L' L''} \braket{Y^{(\text{loc})}_{L} |
  Y^{(\text{glo})}_{L'}} \:
 \bra{Y^{(\text{glo})}_{L'}} 
  G(E) \ket{Y^{(\text{glo})}_{L''}}  \: 
 \nonumber   \\
& & \quad
\times \,   \braket{Y^{(\text{glo})}_{L''} |
  Y^{(\text{loc})}_{L}}
\;\; ,  \nonumber   \\
 & = & 
\sum_{L'} \left|  \braket{Y^{(\text{loc})}_{L} |
  Y^{(\text{glo})}_{L'}} \right|^{2} 
 \nonumber   \\
& & \quad 
\times \, \left( -\frac{1}{\pi} \right) \Im \,
\bra{Y^{(\text{glo})}_{L'}} 
  G(E) \ket{Y^{(\text{glo})}_{L'}}
\;\; ,  \nonumber   \\
 & = & \sum_{L'} U_{L L'} \: n^{(\text{glo})}_{L'}(E)
\;\; .
\label{nloc}
\end{eqnarray}
Specifically in our case we need to describe the situation for
in-plane magnetization, i.e., 
\mm{\bm{M}\|\bm{\hat{x}}}.  The local reference frame is then defined
by the rotation \mm{y\rightarrow y^{\text{(loc)}}}, \mm{z\rightarrow
  x^{\text{(loc)}}}, \mm{x\rightarrow {-z}^{\text{(loc)}}}.
Considering the explicit forms of $Y_{\ell m}^{(\text{glo})}$ and
$Y_{\ell m}^{(\text{loc})}$ for $\ell$=$\ell'$=2, one gets for the $d$
states
\begin{equation}
U_{m m'} \, \equiv \, 
 \left|  \braket{Y^{(\text{loc})}_{2m} |
  Y^{(\text{glo})}_{2m'}} \right|^{2} 
\, = \,
 \left[ 
\begin{array}{rrrrr} 
\frac{1}{16} &\frac{1}{4} & \frac{3}{8} & \frac{1}{4} & \frac{1}{16} 
\\\noalign{\medskip}\frac{1}{4} &\frac{1}{4} & 0 & \frac{1}{4} &
\frac{1}{4} 
\\\noalign{\medskip}\frac{3}{8} & 0 & \frac{1}{4} & 0 &
\frac{3}{8} 
\\\noalign{\medskip}\frac{1}{4} &\frac{1}{4} & 0 & \frac{1}{4} &
\frac{1}{4} 
\\\noalign{\medskip}\frac{1}{16} &\frac{1}{4} & \frac{3}{8} &
\frac{1}{4} & \frac{1}{16}  
\end {array} 
\right] 
\, .
\label{umat}
\end{equation}
More generally, the transformation between the bases is given by
Wigner matrices.\cite{Ros57}

Strictly speaking, Eq.~(\ref{nloc}) with matrix $U$ defined
in~(\ref{umat}) can be used only if the Green function $G$ is diagonal
in the $L$ indices.  This is generally not the case (depending on the
symmetry of the system).  Fortunately, non-diagonal elements of
\mm{\bra{Y^{(\text{glo})}_{L}} G \ket{Y^{(\text{glo})}_{L'}} }\ are
small and can be neglected for the systems we are dealing with.  We
verified this explicitly: If $n^{(\text{loc})}_{L}$ is obtained from
$n^{(\text{glo})}_{L}$ by the transformation (\ref{nloc}), the
$m$-resolved DOS curves obtained thereby agree within the thickness of
the line with curves obtained directly from the Green function via
Eq.~(\ref{locdir}).  It should be noted that this verification ought
to be applied to a sum of the spin components, because for
\mm{\bm{M}\nparallel \bm{\hat{z}}}\ the spin components in
$n^{(\text{glo})}_{L}$ are mixed if they are evaluated directly.
Additionally, the SOC has to be suppressed to get exact equalities.

So far we found a transformation from the global frame to the local
frame.  However, we started our procedure by finding spin-projected
DOS in the local reference frame, so we need an opposite
transformation, from the local frame to the global frame.  A procedure
analogous to that we used to derive Eqs.~(\ref{nloc}) and~(\ref{umat})
cannot be used, because if the Green function is evaluated in the
rotated local reference frame, its non-diagonal elements
\mm{\bra{Y^{(\text{loc})}_{L}} G \ket{Y^{(\text{loc})}_{L'}} }\ cannot
be neglected (the $z^{\text{(loc)}}$ axis of the rotated frame is
chosen in an ``inconvenient'' way --- parallel to the surface).  That
means we have only Eq.~(\ref{nloc}) at our disposal and the
transformation from $n^{(\text{loc})}_{L}$ to $n^{(\text{glo})}_{L}$
has to be accomplished by inverting it.

The inversion of the transformation matrix $U$ defined by
Eq.~(\ref{umat}) cannot be done straightforwardly because this matrix
is singular.  However, the singular 5$\times$5 matrix $U$ of
Eq.~(\ref{umat}) can be folded down to a regular 3$\times$3 matrix
$U^{(\text{fold})}$ if we assume that the $m$-components do not depend
on the sign of $m$, i.e., if
\begin{equation}
n^{(\text{glo})}_{|m|} = n^{(\text{glo})}_{-|m|}
\;\; .
\label{minus}
\end{equation}
In such a case one does not have to deal with all five independent
components $n^{(\text{glo})}_{-2}$, \ldots, $n^{(\text{glo})}_{2}$.
It is enough to keep three of them, $n^{(\text{glo})}_{-2}$,
$n^{(\text{glo})}_{-1}$, $n^{(\text{glo})}_{0}$ and the remaining two
can be recovered by taking
$n^{(\text{glo})}_{2}$=$n^{(\text{glo})}_{-2}$ and
$n^{(\text{glo})}_{1}$=$n^{(\text{glo})}_{-1}$.  This means that the
original Eq.~(\ref{nloc}), which we write here in a more explicit form
as
\[
n^{(\text{loc})}_{m}(E) \: = \: \sum_{m'=-2}^{2} U_{m m'}
n^{(\text{glo})}_{m'}(E) \; , \quad m=-2,\ldots,2 \; ,
\]
is reduced to 
\begin{equation}
n^{(\text{loc})}_{m}(E) \: = \: 
\sum_{m'=-2}^{0} U_{m m'}^{(\text{fold})} n^{(\text{glo})}_{m'}(E)
\; , \quad m=-2,-1,0 
\label{locglo}
\end{equation}
with 
\begin{equation}
U_{m m'}^{(\text{fold})}  \, = \, 
 \left[ 
\begin{array}{rrr} 
\frac{1}{8} &\frac{1}{2} & \frac{3}{8}
\\\noalign{\medskip}\frac{1}{2} &\frac{1}{2} & 0 
\\\noalign{\medskip}\frac{3}{4} & 0 & \frac{1}{4}
\end {array} 
\right] 
\;\; . 
\end{equation}
The matrix $U^{(\text{fold})}$ is regular and can be inverted.  Its
inversion yields a matrix
\begin{equation}
V_{m m'}^{(\text{fold})}  \; = \;
 \left[ 
\begin{array}{rrr} 
-\frac{2}{3} &\frac{2}{3} & 1 
\\\noalign{\medskip}\frac{2}{3} &\frac{4}{3} & -1 
\\\noalign{\medskip}2 & -2 & 1 
\end {array} 
\right] 
\label{vmatfold}
\end{equation}
which transforms the $m$-resolved DOS from the local reference frame
to the global reference frame:
\[
n^{(\text{glo})}_{m}(E) \: = \: 
\sum_{m'=-2}^{0} V_{m m'}^{(\text{fold})} n^{(\text{loc})}_{m'}(E) \; ,
\quad m=-2,-1,0 
\; .
\]
If we assume that the $m$-resolved DOS is independent on the sign of
$m$ not only in the global frame but also in the local frame,
\[
 n^{(\text{loc})}_{|m|} = n^{(\text{loc})}_{-|m|}
\]
[consistently with the fact that the matrix $U$ in Eq.~(\ref{umat}) is
  symmetric], we can unfold the 3$\times$3 matrix $V^{(\text{fold})}$
to a full 5$\times$5 matrix $V$,
\begin{equation}
V_{m m'} \, = \, 
 \left[ 
\begin{array}{rrrrr} 
-\frac{1}{3} &\frac{1}{3} & 1 & \frac{1}{3} & -\frac{1}{3} 
\\\noalign{\medskip}\frac{1}{3} &\frac{2}{3} & -1 & \frac{2}{3} &
\frac{1}{3} 
\\\noalign{\medskip}1 & -1 & 1 & -1 & 1
\\\noalign{\medskip}\frac{1}{3} &\frac{2}{3} & -1 & \frac{2}{3} &
\frac{1}{3} 
\\\noalign{\medskip}-\frac{1}{3} &\frac{1}{3} & 1 & \frac{1}{3} &
-\frac{1}{3}   
\end {array} 
\right] 
\;\; ,
\label{vmat}
\end{equation}
to get a complete transformation of the $m$-resolved DOS from the
local reference frame to the global reference frame:
\begin{equation}
n^{(\text{glo})}_{m}(E) \: = \: 
\sum_{m'=-2}^{2} V_{m m'} n^{(\text{loc})}_{m'}(E) \; , 
\quad m=-2,\ldots,2
\; . 
\label{nglo}
\end{equation}
 
\begin{figure}
\includegraphics[width=8.6cm,viewport=0.2cm 0.3cm 13.0cm 6.0cm]{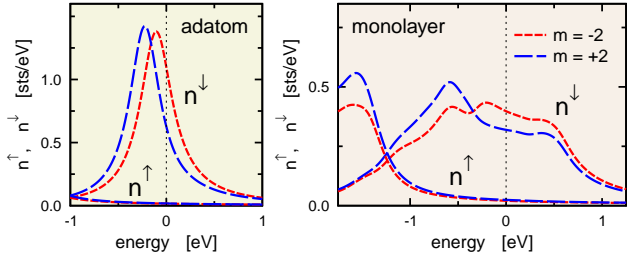}%
\caption{ (Color online) The $m$=-2 and $m$=2 components of the
  majority- and minority-spin DOS for a Co adatom and a Co monolayer
  on Au(111).  Magnetization is out-of-plane.  Data for majority-spin
  DOS are labeled by $n^{\uparrow}$, data for minority-spin DOS by
  $n^{\downarrow}$. }
\label{fig-msplit}
\end{figure}

From the way the transformation Eq.~(\ref{nglo}) was derived it
follows that it can be used only if the $m$-resolved DOS for the
$+|m|$~states is the same as the DOS for the ``$-|m|$~states''.  As a
whole, this is not the case because the SOC splits the
``$\pm|m|$~states''.  Therefore, one cannot simply apply the
transformation (\ref{nglo}) to the minority-spin DOS $m$-resolved in
the local reference frame to get the minority-spin DOS $m$-resolved in
the global reference frame: the minority-spin $\pm|m|$-states are
split by the SOC, therefore the folding of Eq.~(\ref{nloc}) to
Eq.~(\ref{locglo}) cannot be done and the unfolding of the matrix
(\ref{vmatfold}) to a full matrix (\ref{vmat}) cannot be done either.
However, in the energy region we are interested in, i.e., in the
region where the minority-spin states dominate, there is only little
SOC-induced splitting of the $\pm|m|$-states for the {\em
  majority-spin}.  This can be checked explicitly by looking on the
$m$-resolved majority-spin DOS curves in the energy region around
Fermi level.  As an example, we show here the $m$=-2 and $m$=+2 DOS
components for a Co adatom and a Co monolayer (Fig.~\ref{fig-msplit}).
We select the case for \mm{\bm{M}\| \bm{\hat{z}}}, where the
SOC-induced splitting is the largest.  One can see that indeed the
splitting of the majority-spin DOS (labeled by $n^{\uparrow}$) around
\EF\ is much less than the splitting of the minority-spin DOS (labeled
by $n^{\downarrow}$).  So even though the transformation (\ref{nglo})
cannot be applied to the minority-spin DOS, it can be applied to the
majority-spin DOS.

The $m$-resolved minority-spin DOS in the global reference frame can
thus be recovered in the following way: First, let us evaluate the
$m$-resolved DOS directly in the global reference frame, as indicated
in Eq.~(\ref{glodir}). Both global spin channels are strongly mixed
for in-plane magnetization, so there is only a very small difference
between ``spin-up'' and ``spin-down'' $m$-resolved DOS components; if
there is no SOC, even this difference disappears.  By adding
contributions from both spin channels, we get a ``total'' $m$-resolved
DOS in a global frame, with spin components unresolved.  In a second
step, we take the $m$-resolved DOS in the local (rotated) reference
frame, keep only its majority-spin component and transform it to the
global reference frame via Eq.~(\ref{nglo}).  This provides us with a
well-defined majority-spin $m$-resolved DOS in the global reference
frame.  Finally, we subtract this majority-spin DOS $m$-resolved in a
global frame from the total $m$-resolved DOS obtained in the first
step.  This leaves us with minority-spin $m$-resolved DOS in a basis
defined in the global reference frame.  This detour (getting
minority-spin DOS by subtracting majority-spin DOS from the total DOS)
provides more accurate values than what would be obtained if the
transformation (\ref{nglo}) was applied to the minority-spin DOS,
because the condition (\ref{minus}) is satisfied much better for the
majority-spin states than for the minority-spin states in the energy
region of interest.

The procedure described in this appendix should be used only for
systems where there is a substantial exchange splitting between the
majority-spin and minority-spin states.  Only then one can neglect the
SOC-induced $\pm|m|$ splitting of the majority-spin states with
respect to the splitting of the minority-spin states (for energies
where the minority-spin DOS is much larger than the majority-spin
DOS).
\new{ As an indicative parameter whether the procedure can be applied
  or not we suggest the ratio between the exchange splitting
  \mbox{$E_{\downarrow}$-$E_{\uparrow}$}\ and the SOC constant $\xi$.
  Using the parameters given in Tab.~\ref{tab-params}, one gets the
  following values for the ($E_{\downarrow}$-$E_{\uparrow}$)/$\xi$
  ratio: 43.2 for the Fe adatom, 23.1 for the Co adatom, and 5.3 for
  the Ni adatom. }
This illustrates why our procedure works nicely for Fe and Co but not so
well for Ni, as acknowledged in Sec.~\ref{sec-dos}.


\section{DOS for \mm{\bm{M}\|\bm{\hat{x}}} resolved in a local
  reference frame}   
\label{sec-locbas}

\begin{figure}
\includegraphics[width=8.6cm,viewport=0.2cm 0.3cm 13.0cm 10.5cm]{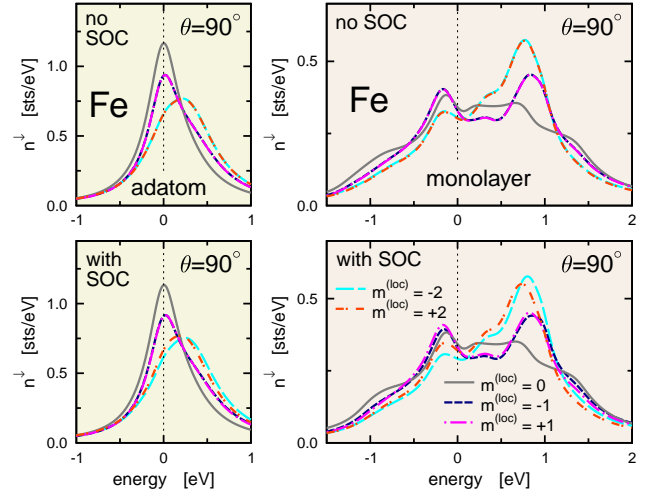}%
\caption{ (Color online) The $d$-component of the minority-spin DOS for a
  Fe adatom (left) and a Fe monolayer (right) on Au(111) for
  magnetization parallel to the surface, with SOC either ignored (top)
  or included (bottom).  The DOS is resolved according to the magnetic
  quantum number $m$ in a local reference frame where the  
  $z^{\text{(loc)}}$ axis is parallel to the Au(111) surface. }
\label{fig-loc90-Fe}
\end{figure}

The DOS presented in Sec.~\ref{sec-dos} was resolved according to the
magnetic quantum number $m$ in a global reference frame, with the $z$
axis perpendicular to the surface.  This ensured the same meaning of
the $m$-components no matter how the magnetization is oriented.
However, one had to apply an additional procedure described in
appendix~\ref{sec-proj} to resolve the spin components.  As this
procedure assumes that SOC does not split the majority-spin DOS which
is not quite the case here (especially for systems with low exchange
splitting such as Ni adatom and monolayer), one might wonder whether
the conclusions based on Figs.~\ref{fig-glob-Fe}--\ref{fig-glob-Ni}
can be trusted.
 
\begin{figure}
\includegraphics[width=8.6cm,viewport=0.2cm 0.3cm 13.0cm 10.5cm]{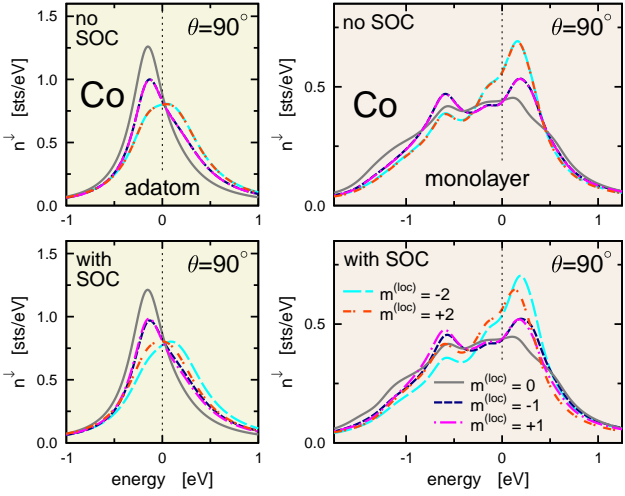}%
\caption{ (Color online) The $d$-component of the minority-spin DOS for a
  Co adatom (left) and a Co monolayer (right) on Au(111) for
  magnetization parallel to the surface.  This figure is analogous to
  Fig.~\protect\ref{fig-loc90-Fe}. }
\label{fig-loc90-Co}
\end{figure}
 
\begin{figure}
\includegraphics[width=8.6cm,viewport=0.2cm 0.3cm 13.0cm 10.5cm]{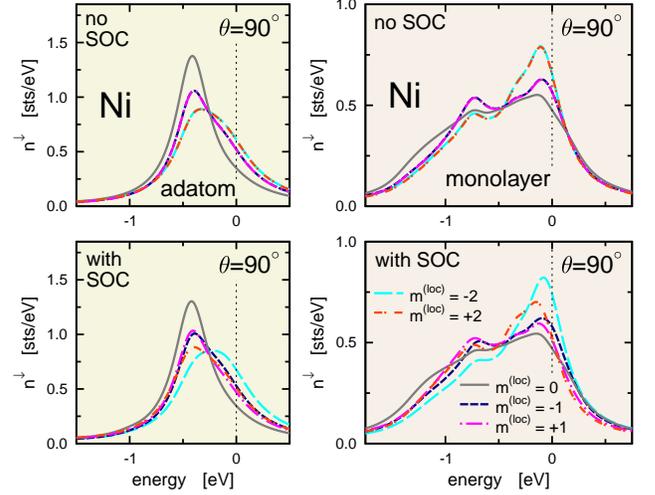}%
\caption{ (Color online) The $d$-component of the minority-spin DOS for a
  Ni adatom (left) and a Ni monolayer (right) on Au(111) for
  magnetization parallel to the surface.  This figure is analogous to
  Fig.~\protect\ref{fig-loc90-Fe}. }
\label{fig-loc90-Ni}
\end{figure}

Therefore, we present in this appendix the $m$-resolved DOS for
$\bm{M}\|\bm{\hat{x}}$ where the magnetic quantum number $m$ refers to
a local reference frame, with the spin quantization axis
$z^{\text{(loc)}}$ parallel to $\bm{M}$.  The outcome is presented in
Figs.~\ref{fig-loc90-Fe}--\ref{fig-loc90-Ni}.  
\new{ Analogous plots for $\bm{M}\|\bm{\hat{z}}$ would be the same as
  respective panels in Figs.~\ref{fig-glob-Fe}--\ref{fig-glob-Ni},
  because in such a case the local and global reference frames
  coincide. }
Note that the
individual $m$-components 
\new{ presented in Figs.~\ref{fig-loc90-Fe}--\ref{fig-loc90-Ni} }
cannot be directly compared to analogous
components in Figs.~\ref{fig-glob-Fe}--\ref{fig-glob-Ni} because their
definitions differ.  This can be clearly seen when comparing the DOS
for systems without SOC, when there can be in principle no
dependence on the magnetization direction.
\new{ The graphs in the second from the top panels of
  Figs.~\ref{fig-glob-Fe}--\ref{fig-glob-Ni} and in the top panels of
  Figs.~\ref{fig-loc90-Fe}--\ref{fig-loc90-Ni} describe the same
  physical situation and yet the individual curves differ }
--- because the
magnetic quantum numbers are defined with respect to different
reference frames.

Even though the $m$-components are defined differently, one can still
qualitatively compare how SOC affects the DOS for $\theta$=0$^{\circ}$
and for $\theta$=90$^{\circ}$.  Concerning the situation for
$\theta$=0$^{\circ}$, one monitors in
Figs.~\ref{fig-glob-Fe}--\ref{fig-glob-Ni} how the plot without SOC
changes if SOC is switched on (i.e., one looks at the two middle
panels of the corresponding figure).  Concerning the situation for
$\theta$=90$^{\circ}$, one monitors analogous changes in
Figs.~\ref{fig-loc90-Fe}--\ref{fig-loc90-Ni}.  One can clearly see
that the effect of SOC is much more pronounced for
$\theta$=0$^{\circ}$ than for $\theta$=90$^{\circ}$.  This confirms
and strengthens the conclusions drawn in Sec.~\ref{sec-dos}, where the
emphasis was put rather on maintaining the possibility for a
component-by-component comparison than on formal correctness.


\bibliography{liter_mech-of-MAE}

\end{document}